\documentclass[%
 aip,
 amsmath,amssymb,
 reprint,%
]{revtex4-1}

\usepackage{graphicx}
\usepackage{dcolumn}
\usepackage{bm}

\usepackage[utf8]{inputenc}
\usepackage[T1]{fontenc}
\usepackage{mathptmx}
\usepackage{etoolbox}

\makeatletter
\def\@email#1#2{%
 \endgroup
 \patchcmd{\titleblock@produce}
  {\frontmatter@RRAPformat}
  {\frontmatter@RRAPformat{\produce@RRAP{*#1\href{mailto:#2}{#2}}}\frontmatter@RRAPformat}
  {}{}
}%
\makeatother
\begin{document}

\preprint{AIP/123-QED}

\title[Tunable quantum logic gate on photonic qubits with a ladder emitter]{Tunable quantum logic gate on photonic qubits with a ladder emitter}
\author{Derek S. Wang}%
\email{derekwang@g.harvard.edu}
\affiliation{ Harvard John A. Paulson School of Engineering and Applied Sciences, Harvard University, Cambridge, MA 02138, USA}
\author{David D. Dai}
\affiliation{The Harker School, San Jose, CA 95129, USA}
\author{Prineha Narang}
\affiliation{Harvard John A. Paulson School of Engineering and Applied Sciences, Harvard University, Cambridge, MA 02138, USA}

\date{\today}

\begin{abstract}
We describe how a ladder emitter can implement a tunable quantum logic gate on photonic qubits encoded in the frequency basis. The ground-to-first excited state of the ladder emitter interacts with the control photon, and the first-to-second excited state transition interacts with the target photon. By controlling the relative detuning between the target photon and the first-to-second excited state transition of the ladder emitter, we enable any controlled-phase operation from 0 to $\pi$. We derive analytical formulas for the performance of the gate through the $S$-matrix formalism, as well as describe the mechanism intuitively. This gate is deterministic, does not utilize any active control, and needs only a single ladder emitter, enabling low-footprint and more efficient decomposition of quantum circuits, especially the quantum Fourier transform. We suggest multiple potential systems for physical realization of our proposal, such as lanthanide ions embedded in Purcell-enhanced cavities. We expect these results to motivate further interest in photonic quantum information processing with designer emitters.
\end{abstract}

\maketitle

\section{Introduction}
Photonic qubits are promising candidates for enabling a universal quantum computer capable of effective integration with long-distance quantum networks because photons offer long coherence times compared to matter based qubits \cite{Slussarenko2019}, qubit transmission at light-speed \cite{Slussarenko2019}, and trivial realization of single-qubit gates via linear-optical components \cite{Fredkin1989}. Due to weak photon-photon interactions without high-quality but difficult-to-fabricate cavities \cite{Li2020, Krastanov2020, Heuck2019, Heuck2020, Zou2020}, the primary challenge in photonic quantum computing is realizing multi-qubit gates \cite{Fredkin1989}. Linear optical quantum computation (LOQC) has provided a partial workaround by using measurement operations to create effective photon-photon interactions. However, LOQC requires a high resource overhead via pre-prepared ancillary photons and is non-deterministic \cite{Knill2001, OBrien2007}. To create a deterministic multi-photonic qubit gate, multiple schemes using mediating systems to create effective photon-photon interactions, such as atoms in a cavities, have been proposed. Many such proposals require active control, such as an external control laser to manipulate the state of a mediating atom, or a microwave field to control a cloud of atoms; these proposals include gates based on Rydberg atoms and the Duan-Kimble proposal and its variants \cite{Gorshkov2011, Tiarks2016, Wu2010, Maller2015, Das2016, Kimble, Koshino2010, Hacker2016}. While some photon-photon gates do not require active control, \textit{i.e.} they are are passive, their footprint is large, such as gates that require an array of many interaction sites \cite{Konyk2019}.

In this \textit{Letter}, we present a scheme for implementing a deterministic, passive, and low-footprint controlled-variable phase gate on photonic qubits. The physical system used is a three-level ladder emitter coupled to a one-dimensional photon field in the reflection geometry, and we encode the qubits as single-photon pulses in the frequency basis. By adjusting the frequencies of the photons relative to the emitter's transition frequencies, we enable any controlled-phase from $0$ to $\pi$. This scheme does not utilize any active control and needs only a single ladder emitter, which may be an advantage for more efficient implementations of multi-qubit gates in quantum circuits for photonic quantum computers.

\section{Gate mechanism}\label{Gate mechanism}

\begin{figure*}[tbhp]
\centering
\includegraphics[width=0.8\textwidth]{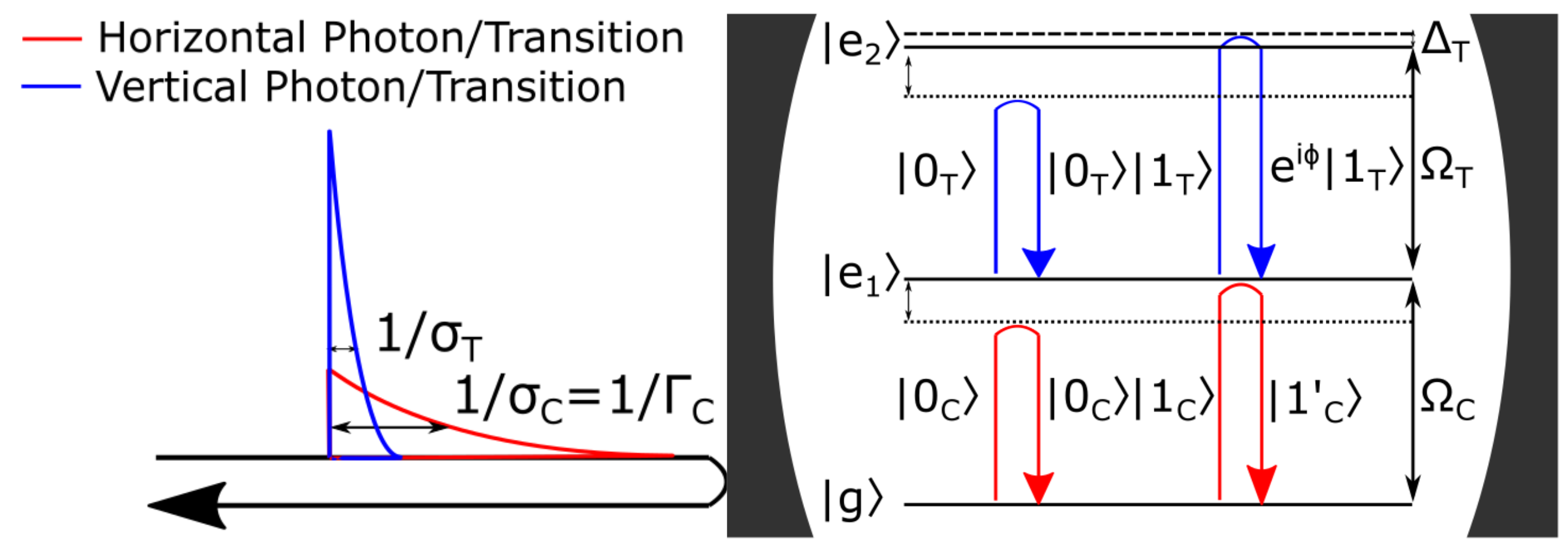}
\caption{Gate schematic. The control photon and the control transition $|{\rm g}\rangle\leftrightarrow|{\rm e_1}\rangle$ of the ladder system are horizontally polarized, and the target photon and target transition $|{\rm e_1}\rangle\leftrightarrow|{\rm e_2}\rangle$ are vertically polarized. The detuning $\Delta_{\rm T}$ of the central frequency of $|1_{\rm T}\rangle$ relative to $\Omega_{\rm T}$ controls the controlled-phase $\phi$ imparted on the target photon.}
\label{fig:schematic}
\end{figure*}

The implementation of the controlled-variable phase gate considered in this \textit{Letter} is illustrated in Fig. \ref{fig:schematic}. The energy level diagram of the emitter comprises a single ground state $|{\rm g}\rangle$, singly excited state $|{\rm e}_1\rangle$, and doubly excited state $|{\rm e}_2\rangle$. The control transition $|\rm{g}\rangle\leftrightarrow|\rm{e_1}\rangle$ has frequency $\Omega_{\rm C}$ and radiative decay rate $\Gamma_{\rm C}$, and we assume it couples to horizontally polarized light due to polarization selection rules. The target transition $|\rm{e_1}\rangle\leftrightarrow|\rm{e_2}\rangle$ has frequency $\Omega_{\rm T}$ and radiative decay rate $\Gamma_{\rm T}$, and it couples to vertically polarized light. The Hamiltonian, setting $\hbar = c = 1$ and assuming the rotating wave approximation, is
\begin{equation}
\begin{split}
    \hat{H}=&\Omega_{\rm{C}}\hat{\sigma}_{\rm{11}}+(\Omega_{\rm{C}}+\Omega_{\rm{T}})\hat{\sigma}_{\rm{22}}\\
    &+\int {\rm d}k_1  \bigg[k_1 \hat{h}^{\dagger}_{k_1}\hat{h}_{k_1} + i\sqrt{\frac{\Gamma_{\rm{C}}}{2\pi}}(\sigma_{\rm{10}}\hat{h}_{k_1} - \sigma_{\rm{01}}\hat{h}^{\dagger}_{k_1})\bigg]\\
    &+\int {\rm d}k_2  \bigg[k_2 \hat{v}^{\dagger}_{k_2}\hat{v}_{k_2} + i\sqrt{\frac{\Gamma_{\rm{T}}}{2\pi}}(\sigma_{\rm{21}}\hat{v}_{k_2} - \sigma_{\rm{12}}\hat{v}^{\dagger}_{k_2})\bigg],
\end{split}
\end{equation}
where $\sigma_{ij}=|{\rm e}_i\rangle\langle {\rm e}_j|$ is the atomic transition operator ($|{\rm e}_0\rangle \equiv |{\rm g}\rangle$), and $\hat{h}_k$ ($\hat{v}_k$) is the annihilation operator for the horizontal (vertical) mode $k$. We have implicitly assumed that the field-atom coupling is constant since the frequency spread of each single-photon packet is small, and only photons resonant or near-resonant with the transitions interact with the system. Because the emitter is in its ground state and there is a single horizontal and a single vertical photon at the start of gate operation, the general photon-emitter wavefunction at all time $t$ is
\begin{equation} \label{eq:eom}
\begin{split}
    |\psi(t)\rangle=&\int {\rm d}k_1 {\rm d}k_2\bigg[f(k_1,k_2,t)e^{-ik_1 t-ik_2 t}\hat{h}^{\dagger}_{k_1}\hat{v}^{\dagger}_{k_2}|0\rangle|\rm{g}\rangle\bigg]\\
    &+\linebreak \int {\rm d}k_2 \bigg[g(k_2,t)e^{-i\Omega_{\rm{C}} t -ik_2 t}\hat{v}^{\dagger}_{k_2}|0\rangle|\rm{e_1}\rangle\bigg]\\
    &+s(t)e^{-i\Omega_{\rm{C}} t -i\Omega_{\rm{T}} t}|0\rangle|\rm{e_2}\rangle.
\end{split}
\end{equation}

We insert this \textit{ansatz} into the time-dependent Schr\"odinger equation to extract the equations of motion for the coefficients in Eq. \eqref{eq:eom}. In addition, at the beginning and end of gate operation, the emitter is in its ground state, and there are two photons. Therefore, we apply the following boundary conditions: only $f(k_1,k_2,\pm\infty)$ is nonzero, and $g(k,\pm\infty)$ and $s(\pm\infty)$ are zero. Integrating the equations of motion from $t=-\infty$ to $t=+\infty$ (see the Supplementary Information for details), the output two-photon wave packet $f(k_1,k_2,+\infty)\equiv f_{\rm out}(k_1,k_2)$ is related to the input two-photon wave packet $f(k_1,k_2,-\infty)\equiv f_{\rm in}(k_1,k_2)$ by
\begin{equation}\label{S_Matrix_1a}
\begin{split}
    &f_{\rm out}(k_1,k_2)=-\frac{\frac{\Gamma_{\rm{C}}}{2}+i(k_1-\Omega_{\rm{C}})}{\frac{\Gamma_{\rm{C}}}{2}-i(k_1-\Omega_{\rm{C}})}f_{\rm{in}}(k_1,k_2)\\
    &+\frac{\Gamma_{\rm{T}}\Gamma_{\rm{C}}/2\pi\int_{-\infty}^{+\infty}d\nu\left[\frac{f_{\rm in}(k_1+\nu,k_2-\nu)}{\frac{\Gamma_{\rm{C}}}{2}-i(k_1+\nu-\Omega_{\rm{C}})}\right]}{\left[\frac{\Gamma_{\rm{C}}}{2}-i(k_1-\Omega_{\rm{C}})\right]\left[\frac{\Gamma_{\rm{T}}}{2}-i(k_1+k_2-\Omega_{\rm{C}}-\Omega_{\rm{T}})\right]}.
\end{split}
\end{equation}

The logical basis state $|1_{\rm C}\rangle$ ($|1_{\rm T}\rangle$) of the control (target) qubit is assigned to a horizontally-polarized (vertically-polarized) single-photon pulse that is resonant with the control $|{\rm g}\rangle\leftrightarrow|{\rm e}_1\rangle$ (target $|\rm{e}_1\rangle\leftrightarrow|\rm{e_2}\rangle$) transition, while the $|0_i\rangle$ logical basis state is strongly detuned and has an identical lineshape and polarization as its $|1_i\rangle$ counterpart. For the target photon, the central frequency of the logical basis state $|1_{\rm T}\rangle$ can be detuned slightly by $\Delta_{\rm T}$ on the order of $\Gamma_{\rm T}$ to adjust the controlled-phase shift $\phi$. After gate operation, the qubits can be trivially separated with a polarizing beam splitter and rotated into the same polarization, should doing so be necessary for the remaining circuit.

\section{Gate performance}

Before presenting gate performance results based on the full expression in Eq. \eqref{S_Matrix_1a}, we outline the intuition of the gate and make analytical approximations to rationalize the conditions necessary for high fidelity. Generally, if a photon is far off-resonant with a transition, the photon will not interact with the transition and pass unchanged, while if a photon is closely resonant, the photon will interact with the transition and pick up a phase. Because the emitter is initialized in its ground state, the second, vertically-polarized target transition is inaccessible to the target photon unless the first and horizontally-polarized transition is first excited by the control photon. Therefore, the state of the horizontally polarized control photon \textit{controls} whether the target photon can interact with the emitter and pick up a phase shift, yielding a controlled-phase gate. Given this framework, we outline the three conditions for high fidelity.

(i) To ensure that an on-resonant control photon will be fully absorbed by the emitter (\textit{i.e.} the control photon can excite the system to its $|\rm{e_1}\rangle$ state fully), it must have a Lorentzian lineshape with bandwidth $\sigma_{\rm C}=\Gamma_{\rm C}$. Intuitively, we can understand this condition as a consequence of time-reversal symmetry: because an emitter initialized in $|\rm{e_1}\rangle$ will emit a photon with Lorentzian lineshape and bandwidth $\sigma_{\rm C}=\Gamma_{\rm C}$ as it decays to $|\rm{g}\rangle$, the same lineshape and bandwith are required to fully excite the emitter from $|\rm{g}\rangle$ to $|\rm{e_1}\rangle$. While the target photon can have any lineshape that can be optimized for maximal gate fidelity, for consistency we assume it also has Lorentzian lineshape. Therefore, we write the two-photon input wave packet as:
\begin{equation} \label{eq:a_in}
    f_{\rm in}(k_1,k_2)=\frac{\sqrt{\frac{\sigma_{\rm{C}}}{2\pi}}}{\frac{\sigma_{\rm{C}}}{2}+i(k_1-\omega_{\rm{C}})}\frac{\sqrt{\frac{\sigma_{\rm{T}}}{2\pi}}}{\frac{\sigma_{\rm{T}}}{2}+i(k_2-\omega_{\rm{T}})},
\end{equation}
where $\sigma_{\rm C}=\Gamma_{\rm C}$. This packet is normalized as $\int {\rm d}k_1 \int {\rm d}k_2 |f_{\rm in}(k_1,k_2)^2| = 1$ and $\langle 0|\hat{h}_{k_1}\hat{v}_{k_2} \hat{h}^{\dagger}_{k_1'}\hat{v}^{\dagger}_{k_2'}|0 \rangle = \delta(k_1'-k_1)\delta(k_2'-k_2)$. The bandwidth of the single-photon pulse $\sigma_{\rm i}$ is the same for logical basis state $|0_i\rangle$ and $|1_i\rangle$, but $\omega_{i}$ depends on whether the photon is in the $|0_i\rangle$ or $|1_i\rangle$ state.

(ii) To ensure that target photon picks up a uniform phase shift across its entire packet, the $|\rm{e_1}\rangle\leftrightarrow|\rm{e_2}\rangle$ transition must adiabatically follow the target photon pulse. Essentially, the target photon is a weighted superposition of different frequency components, with each frequency component picking up a slightly different phase after interacting with the target transition. For this effect to be negligible, the target photon's pulse length $\sim 1/\sigma_{\rm T}$ must be much greater than the lifetime of $|\rm{e_2}\rangle\sim 1/\Gamma_{\rm T}$, or equivalently, the bandwidth of the target photon packet must be much smaller than $\Gamma_{\rm T}$.

(iii) To ensure that the target photon interacts with the ladder emitter only when control photon has been fully absorbed or, equivalently, when the control transition is fully excited, the pulse length of the target photon $\sim 1/\sigma_{\rm T}$ must be much smaller than the pulse length of the control photon $\sim 1/\sigma_{\rm C}$.

Assuming the above three conditions for optimal gate performance, resulting in $\sigma_{\rm C} = \Gamma_{\rm C} \ll \Gamma_{\rm T} \ll \sigma_{\rm T}$, we plug $f_{\rm in}$ from Eq. \eqref{eq:a_in} into Eq. \eqref{S_Matrix_1a} and approximately evaluate it by pulling the slowly varying part of the integrand corresponding to the target packet out of the integral, yielding
\begin{align}\label{S_Matrix_5}
    f_{\rm out}(k_1,k_2)=&\frac{\frac{\Gamma_{\rm{C}}}{2}+i(k_1-\Omega_{\rm{C}})}{\frac{\Gamma_{\rm{C}}}{2}-i(k_1-\Omega_{\rm{C}})}\bigg[-f_{\rm{in}}(k_1,k_2) \nonumber\\
    &+\frac{2}{1-\frac{i(\omega_{\rm{C}}-\Omega_{\rm{C}})}{\Gamma_{\rm{C}}}}\frac{f_{\rm{in}}(k_1,k_2+\frac{\omega_{\rm{C}}-\Omega_{\rm{C}}}{2})}{1-\frac{i(\omega_{\rm{C}}+\omega_{\rm{T}}-\Omega_{\rm{C}}-\Omega_{\rm{T}})}{\Gamma_{\rm{T}}/2}}\bigg].
\end{align}
With Eq. \eqref{S_Matrix_5}, we confirm our intuition for the gate operation. When the control photon is in the $|\rm{0_C}\rangle$ state, we have $|\omega_{\rm{C}}/\Gamma_T-\Omega_{\rm{C}}/\Gamma_T|\gg 1$ and $|\omega_{\rm{C}}/\Gamma_C-\Omega_{\rm{C}}/\Gamma_C|\gg 1$, reducing Eq. \eqref{S_Matrix_5} to $f_{\rm out}(k_1,k_2)=f_{\rm in}(k_1,k_2)$ and demonstrating that both the control and target photon are left unchanged after interacting with the ladder emitter when the control photon is in its $|0_{\rm C}\rangle$ state, regardless of the state of the target qubit. When the control photon is in state $|1_{\rm C}\rangle$, or nearly resonant with the control transition $|{\rm g}\rangle\leftrightarrow|\rm{e_1}\rangle$ by $\Delta_{\rm T}$, Eq. \eqref{S_Matrix_5} reduces to
\begin{equation}\label{TT1}
\begin{split}
    &f_{\rm out}(k_1,k_2)=\\
    &-\frac{1+\frac{i(k_1-\Omega_{\rm{C}})}{\Gamma_{\rm{C}}/2}}{1-\frac{i(k_1-\Omega_{\rm{C}})}{\Gamma_{\rm{C}}/2}}\bigg[-\frac{1+\frac{i(\omega_{\rm{T}}-\Omega_{\rm{T}})}{\Gamma_{\rm{T}}/2}}{1-\frac{i(\omega_{\rm{T}}-\Omega_{\rm{T}})}{\Gamma_{\rm{T}}/2}}\bigg]f_{\rm{in}}(k_1,k_2).
\end{split}
\end{equation}
The target photon, therefore, collects a conditional phase shift $\phi=\rm{arg}[-(1+\frac{i(\omega_{\rm{T}}-\Omega_{\rm{T}})}{\Gamma_{\rm{T}}/2})/(1-\frac{i(\omega_{\rm{T}}-\Omega_{\rm{T}})}{\Gamma_{\rm{T}}/2})]=\pi + 2\arctan{(\frac{\Delta_{\rm T}}{\Gamma_{\rm T}/2})}$ upon re-emission that is dependent on the target photon's detuning $\Delta_{\rm{T}} = \omega_{\rm{T}}-\Omega_{\rm{T}}$. Meanwhile, regardless of the state of the target photon, the control photon is mirrored and rotated in the complex plane from $|1\rangle_{\rm C} = \int {\rm{d}} k_1 \frac{\sqrt{\Gamma_{\rm{C}}/(2\pi)}}{\Gamma_{\rm{C}}/2+i(k_1-\Omega_{\rm{C}})}\hat{h}^{\dagger}_{k_1}|0\rangle$ to $|1'\rangle_{\rm C} = -\int {\rm d}k_1 \frac{\sqrt{\Gamma_{\rm{C}}/(2\pi)}}{\Gamma_{\rm{C}}/2-i(k_1-\Omega_{\rm{C}})}\hat{h}^{\dagger}_{k_1}|0\rangle$. Because this transform is deterministic and always occurs for the $|1_{\rm C}\rangle$ state, it does not impact the gate's information processing capability; it can be viewed as a re-definition of the logical basis $|1_{\rm C}\rangle$ to $|1_{\rm C}'\rangle$. Furthermore, it can be trivially rectified \textit{via} linear optical components or an additional reflection of the control packet \cite{Science1905, Karpi, Pursley2018, Weber2009}; these methods for controlling the pulse shape of single photons may also be used to use a photon as the target photon during one operation of the proposed gate mechanism and the control photon during another. The overall result of the control and target photons interacting with the ladder system is a passive, low-footprint, and deterministic C-PHASE gate on photonic qubits with tunable phase dependent on detuning $\Delta_{\rm T}$. 

\section{Results and Discussion} \label{sec:results}
\begin{figure}[tbhp]
    \centering
    \includegraphics[width=0.8\linewidth]{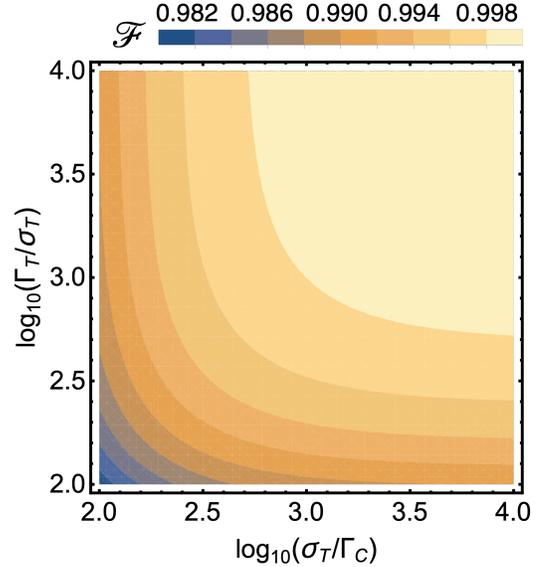}
    \caption{Gate performance. Contour plot of the fidelity $\mathcal{F}$ versus $\Gamma_{\rm{T}}/\sigma_{\rm{T}}$ and $\sigma_{\rm{T}}/\Gamma_{\rm{C}}$ for $\Delta_{\rm T}=0$ and Lorentzian lineshape for the input single-photon pulse of the target qubit.}
    \label{fig:fidelityContour}
\end{figure}

We plug our Lorentzian two-photon input wave packet described in Eq. \eqref{eq:a_in} into Eq. \eqref{S_Matrix_1a} to calculate the fidelity $\mathcal{F}$ and conditional-phase $\phi$ as a function of the photon-emitter system parameters. The fidelity of the gate is defined as $\mathcal{F}=\frac{1}{4}\mathrm{Tr}{(\hat{U}_{\rm ideal}^\dagger \hat{U}_{\rm gate})}$, or equivalently $(\langle{\rm \overline{00}}|\hat{U}_{\rm gate}|{\rm 00}\rangle+\langle{\rm \overline{01}}|\hat{U}_{\rm gate}|{\rm 01}\rangle+\langle{\rm \overline{10}}|\hat{U}_{\rm gate}|{\rm 10}\rangle+\langle{\rm \overline{11}}|\hat{U}_{\rm gate}|{\rm 11}\rangle)/4$, where the state $|\overline{\psi}\rangle$ is the ideal output of the gate given the input state $|\psi\rangle$. By far, the biggest loss to fidelity is caused by the $|\rm{11}\rangle$ input case where both control and target photon interact with the emitter, so the fidelity is essentially $\mathcal{F} = \frac{3}{4}+\frac{1}{4}\langle{\rm \overline{11}}|\hat{U}_{\rm gate}|{\rm 11}\rangle$. The $|\rm{11}\rangle$ input basis state is the only case where nontrivial interaction occurs that may significantly deform the packets of the single-photon pulses. A similar overlap calculation can be done to extract the conditional phase, as detailed in the Supplementary Information, resulting in the following analytic expressions for the fidelity $\mathcal{F}$ and conditional phase $\phi$:
\begin{align} \label{eq:fidelity}
    \mathcal{F} = \frac{3}{4} + \frac{1}{4}|1-\frac{2}{(1+\frac{\Gamma_{\rm C}}{\sigma_{\rm T}})(1+\frac{\sigma_{\rm T}+\Gamma_{\rm C}-2{\rm i} \Delta_{\rm T}}{\Gamma_{\rm T}})}|,
\end{align}
\begin{align} \label{eq:phase}
    \phi = \arg{(1-\frac{2}{(1+\frac{\Gamma_{\rm C}}{\sigma_{\rm T}})(1+\frac{\sigma_{\rm T}+\Gamma_{\rm C}-2{\rm i} \Delta_{\rm T}}{\Gamma_{\rm T}})})}.
\end{align}

In Fig. \ref{fig:fidelityContour}, we explore assumptions (ii) and (iii) by plotting the fidelity $\mathcal{F}$ given by Eq. \eqref{eq:fidelity} for varying $\sigma_{\rm T}/\Gamma_{\rm C}$ and $\Gamma_{\rm T}/\sigma_{\rm T}$ and constant $\Delta_{\rm T}=0$ corresponding to $-\phi \approx \pi$. Fidelity $\mathcal{F}$ increases with both increasing $\sigma_{\rm T}/\Gamma_{\rm C}$ and increasing $\Gamma_{\rm T}/\sigma_{\rm T}$, supporting the intuition that the fidelity improves when the target photon pulse length is smaller than the control photon pulse length and when the bandwidth of the target photon is smaller than the radiative decay rate $\Gamma_{\rm T}$ of the target transition $|{\rm e_2}\rangle\leftrightarrow|{\rm e_1}\rangle$. The fidelity $\mathcal{F}$ is as high as $99.8\%$ for $\sigma_{\rm T}/\Gamma_{\rm C}=\Gamma_{\rm T}/\sigma_{\rm T}=10^3$. Therefore, for a desired gate operation time on the order of 1 $\mu$s, control transition timescales on the order of 1 $\mu$s and target transition timescales on the order of 1 ps are required. 

\begin{figure}[tbhp]
    \centering
    \includegraphics[width=\linewidth]{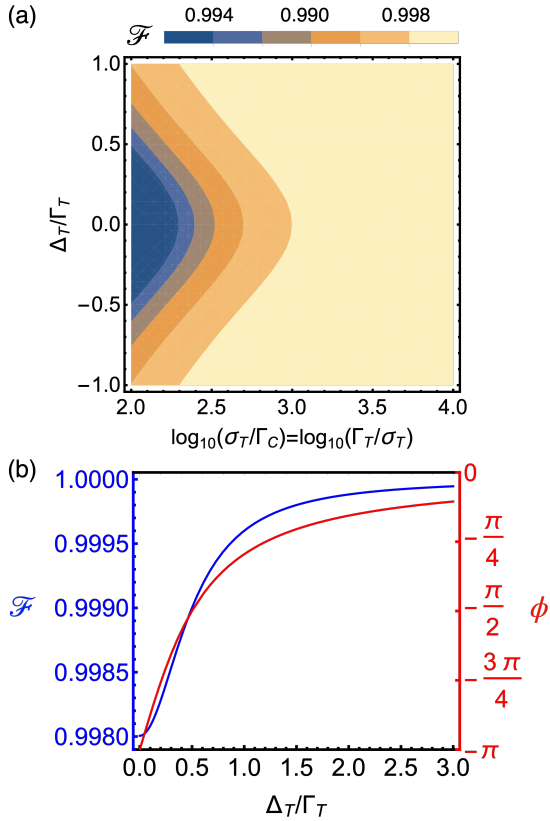}
    \caption{Tunability of the controlled-phase gate. \textbf{(a)} Fidelity $\mathcal{F}$ versus detuning $\Delta_{\rm T}$ and $\sigma_{\rm T}/\Gamma_{\rm C}=\Gamma_{\rm T}/\sigma_{\rm T}$. \textbf{(b)} Phase $\phi$ in radians and fidelity $\mathcal{F}$ versus the detuning $\Delta_{\rm T}$ when $\sigma_{\rm{T}}/\Gamma_{\rm{C}}=\Gamma_{\rm{T}}/\sigma_{\rm{T}}=10^3$. Note that the conditional phase starts near $-\pi$ and goes to zero from the negative side, indicating a phase \textit{lag}.}
    \label{fig:tunability}
\end{figure}

In Fig. \ref{fig:tunability}, we explore how adjusting the detuning $\Delta_{\rm T}$ normalized by $\Gamma_{\rm T}$ impacts the fidelity and phase of the gate. In (a), we see that increasing $\sigma_{\rm T}/\Gamma_{\rm C}$ and $\Gamma_{\rm T}/\sigma_{\rm T}$ increases fidelity, which is consistent with Fig. \ref{fig:fidelityContour}. Additionally, we see that increasing the detuning $\Delta_{\rm T}/\Gamma_{\rm T}$ increases fidelity. In (b), we fix $\sigma_{\rm T}/\Gamma_{\rm C}=\Gamma_{\rm T}/\sigma_{\rm T}=10^3$ and vary $\Delta_{\rm T}/\Gamma_{\rm T}$. Increasing $\Delta_{\rm T}/\Gamma_{\rm T}$ decreases the magnitude of the conditional phase $\phi$ and increases the fidelity $\mathcal{F}$. At $\Delta_{\rm T}/\Gamma_{\rm T}$, $\phi$ is nearly $-\pi$, and $\phi$ approaches $0$ in the $\Delta_{\rm T}/\Gamma_{\rm T}\rightarrow\infty$ limit, indicating a phase lag. The variation of the condition phase $\phi$ with the detuning is consistent with the approximations in the previous section. Enabling continuous phases with magnitudes up to $\pi$ enables greater flexibility when designing quantum circuits, especially those for the quantum Fourier transform that would otherwise require dozens to hundreds of gates to approximate each controlled-variable phase gate using only the CNOT, H, S, and T universal set \cite{Chuang, Kim2018}.

We consider practical implementation based on the constraints of the proposed mechanism for a controlled-variable phase gate on photonic qubits. First, we recall that the difference in central frequencies between $|0_i\rangle$ and $|1_i\rangle$ must be $\gg \Gamma_i$ to ensure that the logical basis states  are well defined. Assuming that $\sigma_{\rm T}/\Gamma_{\rm C}=\Gamma_{\rm T}/\sigma_{\rm T}=10^3$, and that $\sigma_{\rm T}$, $\Gamma_{\rm C}$, and $\Gamma_{\rm T}$ are in the MHz, GHz, and THz range, respectively, for $\mathcal{F}\geq 0.998$, then fulfilling this constraint for $\Omega_i$ in the optical frequency range is trivial. However, photons must lie in the telecommunications band with a bandwidth on the order of a few THz to be seamlessly transmitted without frequency conversion for quantum networking \cite{Vandevender2007, Ding2010, Dreau2018, Maring2018}. In this frequency range, while the difference in central frequencies of the logical basis states of the control photon can be trivially $\gg \Gamma_{\rm C}$, the difference in central frequencies of the logical basis states of the target photon is limited by the bandwidth of the telecommunications band to be of a similar order as $\Gamma_{\rm T}$. As a result, the $|1_{\rm C}\rangle |0_{\rm T}\rangle$ input basis state is particularly susceptible to acquiring extraneous phase and slight wavepacket deformation but still does not appreciably impact fidelity with a deterministic single-qubit phase gate.

A second critical practical consideration is the feasibility of tuning one emitter to apply the proposed gate with variable phases \textit{via} detuning $\Delta_{\rm T}$. As shown in Fig. \ref{fig:tunability}(c), the phase $\phi$ changes with $\Delta_{\rm T}$ on the order of $\Gamma_{\rm T}$, therefore requiring the energy level of the ladder emitter to change on the order of THz to be able to apply controlled-variable phase gates on microsecond timescales. In defect systems, often described as ``artificial atoms" due to their spatial localization and potentially bright, narrow emission, the electric Stark effect has been shown to shift the emission of defect states in transition metal dichalcogenides by up to 5 THz for field strengths of hundreds of MV/m \cite{Chakraborty2017}. While achieving similarly high electric fields to exert similar Stark shifts in conventional atoms or ions may be challenging, it can be more straightforward in Rydberg atoms with larger polarizabilities, resulting in Stark shifts on the order of 0.5 THz for fields of approximately 0.01 MV/m \cite{Petrovic2009}.

A third practical consideration is the validity of the bad-cavity approximation \cite{Turchette1995}, where the spontaneous emission rate of the emitter into free space is assumed to be much smaller than the emission into the waveguide. This regime is the bedrock of other leading proposals for photon-photon gates, including a deterministic $\sqrt{\rm SWAP}$ gate \cite{Koshino2010} and the Duan-Kimble proposal \cite{Kimble} that was experimentally realized \cite{Hacker2016}, where the fidelity was limited by auxiliary optical technologies and not the validity of the bad-cavity approximation. We expect the experimental feasibility of the proposed gate mechanism to be unconstrained by reaching the bad-cavity regime, as it was recently realized in, for instance, several solid-state systems \cite{Peyskens2019, Hausler2020}. 

\section{Conclusion and Outlook}
In summary, we propose a deterministic, passive, and low-footprint controlled variable-phase gate on photonic qubits in the frequency basis using a ladder system to mediate effective photon-photon interactions. Specifically, we analytically derive the scattering matrix for two orthogonally polarized photon pulses interacting with the emitter, and we show that this interaction results in high fidelity for the controlled-variable phase gate given three assumptions of the photon-emitter system. Such a gate enables universal quantum computation when paired with single photon gates, as well as efficient decomposition of fundamental quantum circuits. Furthermore, the ability to encode the target qubit in a different frequency range than the control qubit may enable more facile integration with quantum repeaters in quantum networks or coupling quantum systems operating in different energy ranges to each other \cite{DeGreve2012, Bussieres2014, Saglamyurek2015, Uphoff2016} by averting the need for frequency conversion, where optical photons can be stored locally more conveniently, photons with telecommunication wavelength are more easily transmitted over long distances, and microwave photons can interact with both defect-based spin qubits and superconducting qubits \cite{Neuman2020Nanomagnonics, Wang2020Selection}.

The particular level structure and relative transition rates for the ladder emitter, where $\frac{\sigma_T}{\Gamma_C}=\frac{\Gamma_T}{\sigma_T}\sim10^3$, may be within reach in a variety of physical systems. A particularly interesting potential realization of the ladder emitter are Purcell-enhanced lanthanide ions doped into inert or electro-optical tunable hosts. These lanthanide ions support sub-microsecond long lifetimes, or MHz emission rates, along a ladder of photon-emitting transitions in the optical range that can be Purcell-enhanced \cite{Wu2019} , even dynamically \cite{Xia2021}, although not yet quite at the six orders of magnitude required for the present proposal. Another potential physical system are defect emitters. In Ref. \cite{Su2008}, the authors demonstrate THz emission rates of NV centers in diamond in a Purcell-enhanced optical cavity--this technology in conjunction with silicon vacancy defects in diamond with emission rates nearing the MHz range \cite{Bradac2019} may be on the cusp of enabling experimental realization of the proposed mechanism, although defect emitters that emit photons along an energy ladder have not yet been discovered. First principles-based computational methods with the potential to incorporate the cavity field \cite{flick2017, flick2017ab, Wang2020LossyQEDFT} could also be leveraged to discover new emitters that match the criteria, as has been demonstrated for defects in solid-state materials \cite{Wang2020Hybridized}, although predicting multiply excited states necessary for the three-level ladder emitter remains challenging \cite{Loos2019}. Another method of producing such a ladder emitter is to dipole-couple two three-level emitters, as described in detail in Ref.~\cite{Wang2020}, with the added stipulation that the transition rate from the ground state to one of the excited states is fast, while the transition rate from the ground to the other excited state is slow;  see Appendix \ref{composite} for further details on this added stipulation.

We anticipate that our prediction will spur advances in experimental realizations of passive multi-qubit photonic gates and further searches for candidate emitters to realize the required complex light-matter interactions. In addition, one promising direction for further theoretical studies include coupling the emitters to external fields and sculpted electromagnetic environments to boost the fidelity and improve the practical applicability of the proposed scheme. Another promising direction that warrants further investigation is to generalize the concepts presented in this \textit{Letter} to multi-qubit gates with more control or target qubits, such as the Toffoli gate, as such gates would enable even less resource-intensive decomposition of quantum circuits. We expect these results to motivate further interest in photonic quantum information processing with designer emitters, such as defect complexes in solid-state materials.

\section*{Acknowledgements}
We acknowledge fruitful discussions with Stefan Krastanov, Matthew Trusheim, and Tom\'a\v s Neuman. D.D.D. and D.S.W. contributed equally to this work. This work was supported by the U.S. Department of Energy, Office of Science, Basic Energy Sciences (BES), Materials Sciences and Engineering Division under FWP ERKCK47 `Understanding and Controlling Entangled and Correlated Quantum States in Confined Solid-state
Systems Created via Atomic Scale Manipulation'. D.S.W. is supported by a National Science Foundation Graduate Research Fellowship and partially by the Army Research Office MURI (Ab-Initio Solid-State Quantum Materials) grant number W911NF-18-1-0431. P.N. is a Moore Inventor Fellow and gratefully acknowledges support through Grant GBMF8048 from the Gordon and Betty Moore Foundation.

\section*{Data Availability Statement}

The data that support the findings of this study are available from the corresponding author upon reasonable request.

\appendix
\begin{widetext}
\section{Derivation of General $S$-Matrix} \label{app:derivation}
\noindent
Inserting the \textit{ansatz} of Eq. \eqref{eq:eom} into Schrödinger's equation and multiplying from the left by the appropriate bras yields:
\begin{equation}\label{S-a_mn_TE_1}
    \frac{\mathrm{d}f(k_1,k_2,t)}{\mathrm{d}t}=-\sqrt{\frac{\Gamma_{\rm{C}}}{2\pi}} e^{i(k_1 - \Omega_{\rm{C}})t}g(k_2,t),
\end{equation}
\begin{equation}\label{S-b_n_TE_1}
    \frac{\mathrm{d}g(k_2,t)}{\mathrm{d}t}=+\sqrt{\frac{\Gamma_{\rm{C}}}{2\pi}} \int {\rm d}k_1 e^{-i(k_1- \Omega_{\rm{C}})t}f(k_1,k_2,t)-\sqrt{\frac{\Gamma_{\rm T}}{2\pi}}s(t)e^{i(k_2-\Omega_{\rm T})},
\end{equation}
\begin{equation}\label{S-c_TE_1}
    \frac{\mathrm{d}s(t)}{\mathrm{d}t}=+\sqrt{\frac{\Gamma_{\rm{T}}}{2\pi}} \int {\rm d}k_2  e^{-i(k_2 - \Omega_{\rm{T}})t}g(k_2,t). 
\end{equation}
We derive an analytic formula for the $S$-matrix using repeated formal integrations. First, we integrate Eq. (\ref{S-a_mn_TE_1}) from $\tau=-\infty$ to $\tau=t$:
\begin{equation}\label{S-a_mn_TE_2}
\begin{split}
    f(k_1,k_2,t) - f_{\rm{in}}(k_1,k_2) = -\sqrt{\frac{\Gamma_{\rm{C}}}{2\pi}} \int_{-\infty}^{t}\mathrm{d}\tau\left[e^{i(k_1-\Omega_{\rm{C}})\tau}g(k_2,\tau)\right].
\end{split}
\end{equation}
Plugging Eq. (\ref{S-a_mn_TE_2}) into Eq. (\ref{S-b_n_TE_1}) yields
\begin{equation}\label{S-b_n_TE_2}
\begin{split}
    \frac{\mathrm{d}g(k_2,t)}{\mathrm{d}t}=&\sqrt{\frac{\Gamma_{\rm{C}}}{2\pi}}\int {\rm d}k_1 e^{-i(k_1-\Omega_{\rm{C}})t}f_{\rm{in}}(k_1,k_2)
    -\frac{\Gamma_{\rm{C}}}{2\pi}\int {\rm d}k_1 \int_{-\infty}^{t}\mathrm{d}\tau\left[e^{-i(k_1-\Omega_{\rm{C}})(t-\tau)}g(k_2,\tau)\right]
    -\sqrt{\frac{\Gamma_{\rm{T}}}{2\pi}} e^{i(k_2-\Omega_{\rm{T}})t}s(t).
\end{split}
\end{equation}
By exchanging the integral over modes $k_1$ with the integral over time $t$, we can simplify the second term in Eq. (\ref{S-b_n_TE_2}) to $\frac{\Gamma_{\rm C}}{2}g(k_2,t)$. Moving this term to the left-hand side, multiplying by $e^{\frac{\Gamma_{\rm{C}} t}{2}}$, and applying the product rule yields
\begin{equation}\label{S-b_n_TE_3}
\begin{split}
    \frac{\mathrm{d}}{\mathrm{d}t}\left(g(k_2,t)e^{\frac{\Gamma_{\rm{C}} t}{2}}\right)=& \sqrt{\frac{\Gamma_{\rm{C}}}{2\pi}}\int {\rm d}k_1 e^{\left[\frac{\Gamma_{\rm{C}}}{2}-i(k_1-\Omega_{\rm{C}})\right]t}f_{\rm{in}}(k_1,k_2)
    -\sqrt{\frac{\Gamma_{\rm{T}}}{2\pi}} e^{\left[\frac{\Gamma_{\rm{C}}}{2}+i(k_2-\Omega_{\rm{T}})\right]t}s(t).
\end{split}
\end{equation}
Integrating Eq. \eqref{S-b_n_TE_3} from $-\infty$ to $t$ with the initial condition that $g(k_2,-\infty)=0$ (atom is in its ground state at the start of gate operation) yields
\begin{equation}\label{S-b_n_IF_1}
\begin{split}
    g(k_2,t)=&\sqrt{\frac{\Gamma_{\rm{C}}}{2\pi}}\int {\rm d}k_1 \frac{e^{-i(k_1-\Omega_{\rm{C}})t}f_{\rm{in}}(k_1,k_2)}{\frac{\Gamma_{\rm{C}}}{2}-i(k_1-\Omega_{\rm{C}})}
    -\sqrt{\frac{\Gamma_{\rm{T}}}{2\pi}} e^{-\frac{\Gamma_{\rm{C}}}{2}t}\int_{-\infty}^{t}\mathrm{d}\tau[e^{[\frac{\Gamma_{\rm{C}}}{2}+i(k_2-\Omega_{\rm{T}})]\tau}s(\tau)].
\end{split}
\end{equation}
Inserting Eq. (\ref{S-b_n_IF_1}) into Eq. (\ref{S-c_TE_1}) yields
\begin{equation}\label{S-c_EF_1}
\begin{split}
    \frac{\mathrm{d}s(t)}{\mathrm{d}t}=&\frac{\sqrt{\Gamma_{\rm C}\Gamma_{\rm T}}}{2\pi}\int {\rm d}k_1 {\rm d}k_2\frac{e^{-i(k_1+k_2-\Omega_{\rm{C}}-\Omega_{\rm{T}})t}f_{\rm{in}}(k_1,k_2)}{\frac{\Gamma_{\rm{C}}}{2}-i(k_1-\Omega_{\rm{C}})}
    -\frac{\Gamma_{\rm{T}}}{2\pi}\int {\rm d}k_2 \int_{-\infty}^{t}\mathrm{d}\tau\left[e^{-\left[\frac{\Gamma_{\rm{C}}}{2}+i(k_2-\Omega_{\rm{T}}\right](t-\tau)}s(\tau)\right].
\end{split}
\end{equation}
By exchanging integral over modes and time integral, we can simplify the second term in Eq. \eqref{S-c_EF_1} to $\frac{\Gamma_{\rm{T}}}{2}s(t)$. This manipulation is equivalent to the Markov approximation, as in the derivation of the Wigner-Weisskopf theory \cite{Scully1997}.

Moving this term to the left-hand side, multiplying by $e^{\frac{\Gamma_{\rm{T}}t}{2}}$, and applying the product rule yields
\begin{equation}\label{S-c_EF_2}
\begin{split}
    &\frac{\mathrm{d}}{\mathrm{d}t}\left(s(t)e^{\frac{\Gamma_{\rm{T}} t}{2}}\right)
    =\frac{\sqrt{\Gamma_{\rm C}\Gamma_{\rm T}}}{2\pi} \int {\rm d}k_1 {\rm d}k_2\frac{e^{\left[\frac{\Gamma_{\rm{T}}}{2}-i(k_1+k_2-\Omega_{\rm{C}}-\Omega_{\rm{T}})\right]t}f_{\rm{in}}(k_1,k_2)}{\frac{\Gamma_{\rm{C}}}{2}-i(k_1-\Omega_{\rm{C}})}.
\end{split}
\end{equation}
Integrating Eq. \eqref{S-c_EF_2} from $-\infty$ to $t$ with the initial condition that $s(-\infty)=0$ yields an explicit formula for $s(t)$:
\begin{equation}\label{S-c_EF_3}
\begin{split}
    s(t)=&\frac{\sqrt{\Gamma_{\rm C}\Gamma_{\rm T}}}{2\pi}\int {\rm d}k_1 {\rm d}k_2\bigg(\frac{e^{-i(k_1+k_2-\Omega_{\rm{C}}-\Omega_{\rm{T}})t}f_{\rm{in}}(k_1,k_2)}{\left[\frac{\Gamma_{\rm{C}}}{2}-i(k_1-\Omega_{\rm{C}})\right]\left[\frac{\Gamma_{\rm{T}}}{2}-i(k_1+k_2-\Omega_{\rm{C}}-\Omega_{\rm{T}})\right]}\bigg).
\end{split}
\end{equation}
To calculate the $S$-matrix, we integrate Eq. \eqref{S-a_mn_TE_1} from $-\infty$ to $+\infty$:
\begin{equation}\label{S-a_mn_INT_1}
\begin{split}
    &f(k_1,k_2,t)\bigg|_{-\infty}^{+\infty}=
    -\sqrt{\frac{\Gamma_{\rm{C}}}{2\pi}}\int_{-\infty}^{+\infty}\mathrm{d}t\left[\frac{\mathrm{d}}{\mathrm{d}t}\left(\frac{e^{-\left[\frac{\Gamma_{\rm{C}}}{2}-i(k_1 - \Omega_{\rm{C}})\right]t}}{-\left[\frac{\Gamma_{\rm{C}}}{2}-i(k_1 - \Omega_{\rm{C}})\right]}\right)g(k_2,t)e^{\frac{\Gamma_{\rm{C}} t}{2}}\right].
\end{split}
\end{equation}
Integrating by parts and noting that the non-integral terms will be zero because of the boundary conditions that $g(k_2,-\infty)=g_(k_2,+\infty)=0$ results in
\begin{equation}\label{S-a_mn_INT_2}
\begin{split}
     &f(k_1,k_2,t)\bigg|_{-\infty}^{+\infty}=-\sqrt{\frac{\Gamma_{\rm{C}}}{2\pi}}\int_{-\infty}^{+\infty}\mathrm{d}t\left[\frac{e^{-\left[\frac{\Gamma_{\rm{C}}}{2}-i(k_1 - \Omega_{\rm{C}})\right]t}}{\frac{\Gamma_{\rm{C}}}{2}-i(k_1 - \Omega_{\rm{C}})}\frac{\mathrm{d}}{\mathrm{d}t}\left(g(k_2,t)e^{\frac{\Gamma_{\rm{C}} t}{2}}\right)\right].
\end{split}
\end{equation}
We insert Eq. (\ref{S-b_n_TE_3}) into Eq. (\ref{S-a_mn_INT_2}) and substitute Eq. (\ref{S-c_EF_3}) for $s(t)$ to determine the $S$-matrix only in terms of initial conditions:
\begin{equation}\label{S-S1}
\begin{split}
    &f(k_1,k_2,t)\bigg|_{-\infty}^{+\infty}=
    -\frac{\Gamma_{\rm{C}}}{2\pi}\frac{\int {\rm d}k_1'\int_{-\infty}^{+\infty}\mathrm{d}t\left[e^{-i(k_1'-k_1)t}f_{\rm in}(k_1',k_2)\right]}{\frac{\Gamma_{\rm{C}}}{2}-i(k_1-\Omega_{\rm{C}})}
    +\frac{\sqrt{\Gamma_{\rm C}\Gamma_{\rm T}}}{2\pi}\frac{\int_{-\infty}^{+\infty}\mathrm{d}t\left[e^{+i(k_1+k_2-\Omega_{\rm{C}}-\Omega_{\rm{T}})t}s(t)\right]}{\frac{\Gamma_{\rm{C}}}{2}-i(k_1-\Omega_{\rm{C}})}.
\end{split}
\end{equation}
We compute Eq. \eqref{S-S1} term by term. Using the definition of the delta function, the first term  reduces to $-\frac{\Gamma_{\rm{C}}}{\frac{\Gamma_{\rm{C}}}{2}-i(k_1-\Omega_{\rm{C}})}f_{\rm{in}}(k_1,k_2)$. Inserting $s(t)$ into the second term yields
\begin{equation}\label{S-S2}
\begin{split}
    &\frac{\Gamma_{\rm C}\Gamma_{\rm T}/(2\pi)^2}{\frac{\Gamma_{\rm{C}}}{2}-i(k_1-\Omega_{\rm{C}})}\int {\rm d}k_1' {\rm d}k_2'
    \bigg(\frac{\int_{-\infty}^{+\infty}\mathrm{d}t\left[e^{-(k_1'+k_2'-k_1-k_2)t}f_{\rm in}(k_1',k_2')\right]}{\left[\frac{\Gamma_{\rm{C}}}{2}-i(k_1'-\Omega_{\rm{C}})\right]\left[\frac{\Gamma_{\rm{T}}}{2}-i(k_1'+k_2'-\Omega_{\rm{C}}-\Omega_{\rm{T}})\right]}\bigg).\\
\end{split}
\end{equation}
Applying the definition of the delta function to the time integral yields:
\begin{equation}\label{S-R3}
\begin{split}
    =\frac{\Gamma_{\rm{T}}\Gamma_{\rm{C}}/(2\pi)^2}{\frac{\Gamma_{\rm{C}}}{2}-i(k_1-\Omega_{\rm{C}})}
    \int_{-\infty}^{+\infty}\mathrm{d}k_1'\int_{-\infty}^{+\infty}\mathrm{d}k_2'\bigg(\frac{f_{\rm{in}}(k_1',k_2')2\pi\delta(k_1'+k_2'-k_1-k_2)}{\left[\frac{\Gamma_{\rm{C}}}{2}-i(k_1'-\Omega_{\rm{C}})\right]\left[\frac{\Gamma_{\rm{T}}}{2}-i(k_1'+k_2'-\Omega_{\rm{C}}-\Omega_{\rm{T}})\right]}\bigg).
\end{split}
\end{equation}
The delta function removes one of the integrals, yielding:
\begin{equation}\label{S-R4}
\begin{split}
    =\frac{\Gamma_{\rm{T}}\Gamma_{\rm{C}}/2\pi}{\left[\frac{\Gamma_{\rm{C}}}{2}-i(k_1-\Omega_{\rm{C}})\right]\left[\frac{\Gamma_{\rm{T}}}{2}-i(k_1+k_2-\Omega_{\rm{C}}-\Omega_{\rm{T}})\right]}\int_{-\infty}^{+\infty}\mathrm{d}\nu\left[\frac{f_{\rm{in}}(k_1+\nu,k_2-\nu)}{\frac{\Gamma_{\rm{C}}}{2}-i(k_1+\nu-\Omega_{\rm{C}})}\right].\\
\end{split}
\end{equation}
Combining both terms yields the complete $S$-matrix:
\begin{equation}\label{S-S_Matrix_1}
\begin{split}
    &f_{\rm{out}}(k_1,k_2)=-\frac{\frac{\Gamma_{\rm{C}}}{2}+i(k_1-\Omega_{\rm{C}})}{\frac{\Gamma_{\rm{C}}}{2}-i(k_1-\Omega_{\rm{C}})}f_{\rm{in}}(k_1,k_2)+\frac{\Gamma_{\rm{T}}\Gamma_{\rm{C}}/2\pi\int_{-\infty}^{+\infty}\mathrm{d}\nu\frac{f_{\rm{in}}(k_1+\nu,k_2-\nu)}{\frac{\Gamma_{\rm{C}}}{2}-i(k_1+\nu-\Omega_{\rm{C}})}}{\left[\frac{\Gamma_{\rm{C}}}{2}-i(k_1-\Omega_{\rm{C}})\right]\left[\frac{\Gamma_{\rm{T}}}{2}-i(k_1+k_2-\Omega_{\rm{C}}-\Omega_{\rm{T}})\right]}.
\end{split}
\end{equation}
\section{Derivation of Gate Performance} \label{app:general}
While the general $S$-matrix in Eq. \eqref{S-S_Matrix_1} can be solved numerically for any input photon packets, we apply the approximations discussed in Section \ref{Gate mechanism} to generate physical intuition for the numerical results.
We first write $f_{\rm{in}}(k_1,k_2) = \frac{\sqrt{\frac{\Gamma_{\rm{C}}}{2\pi}}}{\frac{\Gamma_{\rm{C}}}{2}+i(k_1-\omega_{\rm{C}})}f(k_2)$, where $\omega_{\rm{C}}$ is the control packet's central frequency and $f(\omega)$ is any normalized lineshape satisfying the assumptions of the target photon. According to condition (ii) in Section \ref{Gate mechanism}, $f(k)$ is slowly varying compared to the rest of the integrand in Eq. (\ref{S-S_Matrix_1}), so we may pull it out and evaluate it at the argmax of the remaining quantities in the integrand. Substituting $u = k_1 + v - \Omega_{\rm{C}}$ and detuning $\Delta_C=\omega_{\rm{C}}-\Omega_{\rm{C}}$ and completing the square in the denominator, the integral becomes
\begin{equation}\label{S-S_Matrix_3}
    \sqrt{\frac{\Gamma_{\rm{C}}}{2\pi}}\int_{-\infty}^{+\infty}\mathrm{d}u\frac{1}{(\frac{\Gamma_{\rm{C}}}{2}-\frac{i\Delta_C}{2})^2+(u-\frac{\Delta_C}{2})^2}.
\end{equation}
The integral itself can be determined via arctangent substitution as $\frac{\pi}{\frac{\Gamma_{\rm{C}}}{2}-\frac{i\Delta_C}{2}}$. The integrand takes its maximal value at $u=\frac{\Delta_C}{2}$, so we use $f(k_1+k_2-\frac{\omega_{\rm{C}}+\Omega_{\rm{C}}}{2})$ as the value of the target packet. Thus, Eq. \eqref{S-S_Matrix_1} becomes
\begin{equation}\label{S-S_Matrix_4}
\begin{split}
    &f_{\rm{out}}(k_1,k_2)=-\frac{\frac{\Gamma_{\rm{C}}}{2}+i(k_1-\Omega_{\rm{C}})}{\frac{\Gamma_{\rm{C}}}{2}-i(k_1-\Omega_{\rm{C}})}\bigg[f_{\rm{in}}(k_1,k_2)-\frac{2}{1-\frac{i\Delta_C}{\Gamma_{\rm{C}}}}\frac{f_{\rm{in}}(k_1,k_2+k_1-\frac{\omega_{\rm{C}}+\Omega_{\rm{C}}}{2})}{1-\frac{i(k_1+k_2-\Omega_{\rm{C}}-\Omega_{\rm{T}})}{\Gamma_{\rm{T}}/2}}\bigg].
\end{split}
\end{equation}
When the control is in the $|0_\mathrm{C}\rangle$ state, $k_1-\Omega_{\rm C} \gg \Gamma_{\rm C}$, so the prefactor in front of the square brackets in Eq. \eqref{S-S_Matrix_4} is essentially $1$ and the second term in the square brackets is suppressed. Thus, regardless of the state of the target photon, both the target and control packets are unchanged by the interaction with the ladder emitter: $f_{\rm{out}}(k_1,k_2) = f_{\rm{in}}(k_1,k_2)$. 

Next, we consider the case of a resonant control photon corresponding to the $|1_\mathrm{C}\rangle$ state. Because the bandwidth of the target packet is much wider in frequency than the control packet's, for a resonant control photon, we may approximate $f_{\rm{in}}(k_1,k_2+k_1-\frac{\omega_{\rm{C}}+\Omega_{\rm{C}}}{2})$ as $f_{\rm{in}}(k_1,k_2)$:
\begin{equation}\label{S-TT1}
\begin{split}
    &f_{\rm{out}}(k_1,k_2)=    -\frac{\frac{\Gamma_{\rm{C}}}{2}+i(k_1-\Omega_{\rm{C}})}{\frac{\Gamma_{\rm{C}}}{2}-i(k_1-\Omega_{\rm{C}})}\bigg[-\frac{1+\frac{i(k_1+k_2-\Omega_{\rm{C}}-\Omega_{\rm{T}})}{\Gamma_{\rm{T}}/2}}{1-\frac{i(k_1+k_2-\Omega_{\rm{C}}-\Omega_{\rm{T}})}{\Gamma_{\rm{T}}/2}}\bigg]f_{\rm{in}}(k_1,k_2),
\end{split}
\end{equation}
which is identical to Eq. \eqref{TT1}. Because the bandwidths of the control and target photons are very small compared to $\Gamma_{\rm T}$, we may replace $\frac{i(k_1+k_2-\Omega_{\rm{C}}-\Omega_{\rm{T}})}{\Gamma_{\rm{T}}/2}$ by its mean-value $\frac{i\Delta_{\rm{T}}}{\Gamma_{\rm{T}}/2}$. Eq. \eqref{S-TT1} becomes
\begin{equation}\label{S-TT2}
    f_{\rm{out}}(k_1,k_2)=-\frac{\frac{\Gamma_{\rm{C}}}{2}+i(k_1-\Omega_{\rm{C}})}{\frac{\Gamma_{\rm{C}}}{2}-i(k_1-\Omega_{\rm{C}})}\bigg[-\frac{1+\frac{i\Delta_{\rm{T}}}{\Gamma_{\rm{T}}/2}}{1-\frac{i\Delta_{\rm{T}}}{\Gamma_{\rm{T}}/2}}\bigg]f_{\rm{in}}(k_1,k_2).
\end{equation}
To calculate the variable-phase $\phi$ and fidelity $\mathcal{F}$ analytically when both control and target have Lorentzian lineshape, we evaluate the quantity
\begin{equation}\label{S-overlap_1}
\begin{split}
    &Z = \langle{\rm 1'_C}|\langle{\rm 1_T}|\hat{U}_{\rm gate}|{\rm 1_C}\rangle|{\rm 1_T}\rangle=\int_{-\infty}^{+\infty}\int_{-\infty}^{+\infty}{\rm d}k_1{\rm d}k_2
    \bigg[-\frac{\sqrt{\frac{\Gamma_{\rm{C}}}{2\pi}}}{\frac{\Gamma_{\rm{C}}}{2}+i(k_1-\omega_{\rm{C}})}\frac{\sqrt{\frac{\sigma_{\rm{T}}}{2\pi}}}{\frac{\sigma_{\rm{T}}}{2}-i(k_2-\omega_{\rm{T}})}f_{\rm{out}}(k_1,k_2)\bigg],
\end{split}
\end{equation}
for the case of $\omega_{\rm C} = \Omega_{\rm C}$ and $\omega_{\rm T} - \Omega_{\rm T} = \Delta_{\rm T}$, where both photons interact with the emitter and corresponding to the input states that systemically result in the worst gate performance. As described in the main text, the fidelity $\mathcal{F}$ is $\frac{3}{4}+\frac{1}{4}|Z|$, and the variable phase $\phi$ is $\textrm{arg}(Z)$. We insert our expression for $f_{\rm{out}}(k_1,k_2)$:
\begin{equation}\label{S-Z_1}
    Z = Z_1+Z_2,
\end{equation}
where $Z_1$ is
\begin{equation}\label{S-Z1-1}
\begin{split}
    &Z_1 = \int_{-\infty}^{+\infty}\int_{-\infty}^{+\infty}{\rm d}k_1{\rm d}k_2\bigg[\frac{\frac{\Gamma_{\rm{C}}}{2\pi}}{(\frac{\Gamma_{\rm{C}}}{2})^2+(k_1-\Omega_{\rm{C}})^2}\frac{\frac{\sigma_{\rm{T}}}{2\pi}}{(\frac{\sigma_{\rm{T}}}{2})^2+(k_2-\omega_{\rm{T}})^2}\bigg],
\end{split}
\end{equation}
and $Z_2$ is
\begin{equation}\label{S-Z2-1}
\begin{split}
    &Z_2 = -\frac{\Gamma_{\rm T}\Gamma_{\rm C}^{2}\sigma_{\rm T}}{(2\pi)^2}\int_{-\infty}^{+\infty}\int_{-\infty}^{+\infty}\int_{-\infty}^{+\infty}{\rm d}k_1{\rm d}k_2{\rm d}\nu\bigg[
    \frac{1}{\left[(\frac{\Gamma_{\rm{C}}}{2})^2+(k_1-\Omega_{\rm{C}})^2\right]\left[\frac{\sigma_{\rm{T}}}{2}-i(k_2-\omega_{\rm{T}})\right]}\\
    &\frac{1}{\left[(\frac{\Gamma_{\rm{C}}}{2})^2+(k_1+\nu-\Omega_{\rm{C}})^2\right]\left[\frac{\sigma_{\rm{T}}}{2}+i(k_2-\nu-\omega_{\rm{T}})\right]}
    \frac{1}{\left[\frac{\Gamma_T}{2}-i(k_1+k_2-\Omega_{\rm C}-\Omega_{\rm T})\right]}\bigg].
\end{split}
\end{equation}
Because the photons are normalized, $Z_1$ trivially is $1$. $Z_2$ can be simplified with partial-fraction decomposition to yield
\begin{equation}\label{S-Z2-3}
    Z_2=-\frac{2}{(1+\frac{\Gamma_{\rm C}}{\sigma_{\rm T}})(1+\frac{\Gamma_{\rm C}+\sigma_{\rm T}}{\Gamma_{\rm T}}-i\frac{\Delta_{\rm T}}{\Gamma_{\rm T}/2})},
\end{equation}
thus obtaining the expressions for the fidelity $\mathcal{F}$ and phase $\phi$.

\end{widetext}

\section{Ladder Emitter via Dipole-Coupled Emitters}\label{composite}

We briefly review the construction of a composite emitter from dipole-coupled emitters following Ref. \citenum{Wang2020, wang2021entangled} and then describe how this composite emitter approximates the ladder emitter required for implementation of the proposed gate mechanism.

Consider a system consisting of two three-level systems denoted by $i\in\{\alpha,\beta\}$. Each three-level system consists of a ground state $|g_i\rangle$, excited state $|x_i\rangle$ with energy $\hbar\omega_x$ and transition dipole moment $\bm{d}_{x_i} =\langle x_i | {\rm e}\bm{r} | g_i \rangle=d_{x_i} \hat{x}$, and excited state $|y_i\rangle$ with energy $\hbar\omega_y$ and  transition dipole moment $\bm{d}_{y_i}=\langle y_i | {\rm e}\bm{r} | g_i \rangle=d_{y_i}\hat{y}$, where $\bm{r}$ is the position operator and ${\rm e}$ is the electron charge. The Hamiltonian $H_i$ of each isolated three-level system can therefore be written as $H_i = \hbar\omega_x |x_i \rangle \langle x_i| + \hbar\omega_y | y_i \rangle \langle y_i|$.

When emitters $\alpha$ and $\beta$ at positions $\bm{r}_\alpha$ and $\bm{r}_\beta$, respectively, are brought close and couple via electric dipole interactions, the total electronic Hamiltonian $H_{\rm el}$ can be written in the product space of the two three-level systems as
\begin{align} \label{eq:hamiltonian}
    H_{\rm el} = {H_{\alpha\beta}} + H_{\rm dip}, 
\end{align}
where $H_{\alpha\beta}=H_\alpha+H_\beta$, and the dipole-coupling Hamiltonian $H_{\rm dip}$, in the rotating wave approximation (RWA) where we have dropped double \mbox{(de-)excitations,} is given by
\begin{align}
    H_{\rm dip}=\sum_{pq\in \{x,y\}}J_{pq}(|g p\rangle \langle q g| +|q g\rangle \langle g p|),
\end{align}
where $|rs\rangle\equiv |r_\alpha\rangle |s_\beta\rangle$ with $r,s \in \{g, x, y\}$, and transition dipole moments are real. The dipole interaction energy $J_{pq}$ is \cite{Lukin2000}
\begin{align}
    J_{pq} = \frac{|\bm{d}_{p_\alpha} ||\bm{d}_{q_\beta} |}{4\pi\epsilon_0\epsilon_r|\bm{r}_\alpha-\bm{r}_\beta|^3}\left[\bm{e}_{p_\alpha}\cdot\bm{e}_{q_\beta}-3(\bm{e}_{p_\alpha}\cdot\bm{n})(\bm{e}_{q_\beta}\cdot\bm{n})\right],
\end{align}
where $\epsilon_r$ is the relative permittivity of the host material, $\bm{e}_{s_i}$ is the unit vector of the dipole moment $\bm{d}_{s_i}$, and $\bm{n}$ is the unit vector of $\bm{r}_\alpha-\bm{r}_\beta$.

Assuming for the sake of simplicity that $\bm{n}$ lies on the $x$-axis and that the dipole moments of the same polarizations of emitters $\alpha$ and $\beta$ are identical ($d_x\equiv d_{x_\alpha}=d_{x_{\beta}}$ and $d_y \equiv d_{y_\alpha}=d_{y_{\beta}}$), $H_{\rm el}$ can be diagonalized to produce nine eigenstates with eigenenergies listed in Table \ref{tab:eigen}. Non-zero transition dipole moments between these eigenstates, corresponding to dipole-allowed and photon-emitting transitions, are listed in Table \ref{tab:dipole}. The subscripts ``A" and ``S" stand for ``anti-symmetric" and ``symmetric" combinations, respectively. Notably, direct transitions between symmetric and anti-symmetric states are dipole-forbidden. Assuming that this composite system starts in the ground state $|g\rangle$, population of only the symmetric states is allowed \textit{via} dipole-allowed transitions. 

Importantly, there are two paths to the doubly excited state $|xy_{\rm S}\rangle$ from $|g\rangle$. The first path follows $|g\rangle \leftrightarrow |x_{\rm S}\rangle $ with a transition dipole moment of $\sqrt{2}d_x\hat{x}$ interacting with an $x$-polarized photon with energy $\hbar\omega_x-J_{xx}$ to $|x_{\rm S}\rangle \leftrightarrow |xy_{\rm S}\rangle$ with a transition dipole moment of $d_y\hat{y}$ interacting with a $y$-polarized photon with energy $\hbar\omega_y+J_{xx}$. The second path follows $|g\rangle \leftrightarrow |y_{\rm S}\rangle $ with a transition dipole moment of $\sqrt{2}d_y\hat{y}$ interacting with a $y$-polarized photon with energy $\hbar\omega_y+J_{yy}$ to $|y_{\rm S}\rangle \leftrightarrow |xy_{\rm S}\rangle$ with a transition dipole moment of $d_x\hat{x}$ interacting with an $x$-polarized photon with energy $\hbar\omega_x-J_{yy}$. 

To approximate the composite system of dipole-coupled emitters as the desired composite system, we simply assume that $d_x \ll d_y$, assign $|x_{\rm S}\rangle\equiv |e_1\rangle$ and $|xy_{\rm S}\rangle\equiv |e_2\rangle$ as shown in Fig. \ref{fig:schematic}, and set the energy of the $|1\rangle$ state and polarization of the control (target) photon to $\hbar\omega_x-J_{xx}$ ($\hbar\omega_y+J_{xx}$) and $x$- ($y$-) polarization, respectively. Setting $d_x \ll d_y$ ensures that the $|g\rangle\leftrightarrow|e_1\rangle$ control transition of the first pathway is relatively slow and that the $|x_{\rm S}\rangle \leftrightarrow |e_2\rangle$ target transition of the first pathway is relatively fast, assuming Fermi's Golden Rule-like scaling of the emission rate with the transition dipole moment. The composite system cannot be excited to $|e_2\rangle$ from $|g\rangle$ due to the polarization and frequency requirements of the second pathway, and during de-excitation from $|e_2\rangle$ to $|g\rangle$, the first pathway is dominant because de-excitation \textit{via} the second pathway from $|e_2\rangle$ to $|y_{\rm S}\rangle$ is relatively slow compared to the de-excitation from $|e_2\rangle$ to $|x_{\rm S}\rangle$, again based on their transition dipole moments. Therefore, we see that the composite system of dipole-coupled emitters can be approximated as the desired composite system, although further detailed numerical study to determine the impact of the second pathway on the fidelity of the proposed gate operation.

\begin{table}[h!]
\centering
\begin{tabular}{ ||c|c|c||} 
\hline
& Eigenstate & Eigenenergy \\
 \hline
1 & $|g\rangle \equiv |g g\rangle$ & $\hbar\omega_g=0$ \\
2 & $|y_{\rm A}\rangle \equiv \frac{1}{\sqrt{2}}(|gy\rangle - |yg\rangle)$ & $\hbar\omega_{y_{\rm A}}=\hbar\omega_y-J_{yy}$ \\
3 & $|y_{\rm S}\rangle \equiv \frac{1}{\sqrt{2}}(|gy\rangle + |yg\rangle)$ & $\hbar\omega_{y_{\rm S}}=\hbar\omega_y+J_{yy}$\\
4 & $|x_{\rm S}\rangle \equiv \frac{1}{\sqrt{2}}(|gx\rangle + |xg\rangle)$ & $\hbar\omega_{x_{\rm S}}=\hbar\omega_x-J_{xx}$\\
5 & $|x_{\rm A}\rangle \equiv \frac{1}{\sqrt{2}}(|gx\rangle - |xg\rangle$ & $\hbar\omega_{x_{\rm A}}=\hbar\omega_x+J_{xx}$\\
6 & $ |yy\rangle$ & $\hbar\omega_{yy}=2\hbar\omega_y$\\
7 & $|xy_{\rm S}\rangle \equiv \frac{1}{\sqrt{2}}(|xy\rangle + |yx\rangle)$ & $\hbar\omega_{xy_{\rm S}}=\hbar(\omega_x+\omega_y)$\\ 
8 & $|xy_{\rm A}\rangle \equiv \frac{1}{\sqrt{2}}(|xy\rangle - |yx\rangle)$ & $\hbar\omega_{xy_{\rm A}}=\hbar\omega_{xy_{\rm S}}$\\
9 & $|xx\rangle$ & $\hbar\omega_{xx}=2\hbar\omega_{x}$ \\
\hline
\end{tabular}
\caption{Eigenstates and eigenenergies of $H_{\rm el}$. Reproduced from Ref. \cite{Wang2020}.}
\label{tab:eigen}
\end{table}

\begin{table}[h!]
\centering
\begin{tabular}{ ||c|c|c||} 
\hline
Initial & Final & $\bm{d}$ \\
 \hline
$|g\rangle$ & $|x_{\rm S} \rangle$ & $\sqrt{2}d_x\hat{x}$ \\
$|g\rangle$ & $|y_{\rm S} \rangle$ & $\sqrt{2}d_y\hat{y}$ \\
$|x_{\rm S}\rangle$ & $|xy_{\rm S}\rangle$ & $d_y\hat{y}$ \\
$|y_{\rm S} \rangle$ & $|xy_{\rm S}\rangle$ & $d_x\hat{x}$ \\
$|x_{\rm S} \rangle$ & $|xx\rangle$ & $\sqrt{2} d_x\hat{x}$ \\
$|y_{\rm S} \rangle$ & $|yy\rangle$ & $\sqrt{2} d_y\hat{y}$ \\
$|x_{\rm A}\rangle$ & $|xy_{\rm A}\rangle$ & $d_y\hat{y}$ \\
$|y_{\rm A}\rangle$ & $|xy_{\rm A}\rangle$ & $d_x\hat{x}$ \\
\hline
\end{tabular}
\caption{The dipole operator $\bm{d}$ in the eigenbasis. Reproduced from Ref. \cite{Wang2020}.}
\label{tab:dipole}
\end{table}

\newcommand{\noopsort}[1]{} \newcommand{\printfirst}[2]{#1}
  \newcommand{\singleletter}[1]{#1} \newcommand{\switchargs}[2]{#2#1}

\end{document}